\documentclass[final,1p,times]{elsarticle} 

\usepackage{graphicx}
\usepackage{amssymb} 
\usepackage{amsthm} 
\usepackage{lineno}

\usepackage{hyperref}

\usepackage{sidecap}

 \biboptions{sort&compress}

\newcommand{\pt}{p_{\rm T}}

\newcommand{\pp}{pp}

\newcommand{\sqrts}{\sqrt{s}}
\newcommand{\sqrtsnn}{\sqrt{s_{\rm NN}}}

\newcommand{\mev}{\mathrm{MeV}}
\newcommand{\gev}{\mathrm{GeV}}

\newcommand{\PbPb}{\mbox{Pb--Pb}}

\newcommand{\Jpsi}{{\rm J}/\psi}

\newcommand{\Dzero}{{\rm D^0}}

\newcommand{\Dstar}{{\rm D^{*+}}}

\newcommand{\Dplus}{{\rm D^+}}

\newcommand{\Ds}{{\rm D_{\rm s}^{+}}}

\newcommand{\RAA}{R_{\rm AA}}
\newcommand{\TAA}{T_{\rm AA}}
\newcommand{\vtwo}{v_{2}}

\journal{Nuclear Physics A} 

\begin{document}

\begin{frontmatter} 

\title{Heavy-flavor suppression and azimuthal anisotropy in \PbPb~collisions at $\sqrtsnn=2.76$~TeV
	with the ALICE detector}

\author{Zaida Conesa del Valle (for the ALICE\fnref{col1} Collaboration)}
\fntext[colALICE] {A list of members of the ALICE Collaboration and acknowledgements can be found at the end of this issue.}
\address{European Organization for Nuclear Research (CERN), Geneva, Switzerland \\
	Institut Pluridisciplinaire Hubert Curien (IPHC), Universit\'e deStrasbourg, CNRS-IN2P3, Strasbourg, France}

\begin{abstract} 
Heavy-flavor hadron production studies in \PbPb~collisions at $\sqrtsnn=2.76$~TeV with the 2010 and 2011 data samples are presented. The measurements are performed with the ALICE detector in various decay channels and in a wide kinematic range. 
Heavy-flavor hadrons exhibit a suppression in the most central \PbPb~collisions that amounts to a factor 3--5 for $\pt \sim 8$--$10~\gev/c$. The second harmonic of the azimuthal distribution Fourier decomposition, $\vtwo$, is non-zero for non-central collisions at intermediate $\pt$ ($\sim3~\gev/c$).
\end{abstract} 

\end{frontmatter} 


\section{Why studying heavy-flavor hadron production?}

Charm and beauty quarks are produced in the early stage of the collision. Their production in proton--proton (\pp)~collisions at the LHC is a tool to test pQCD calculations in a new energy domain. 
Their spectra in heavy-ion interactions are influenced by the formation of hot and dense QCD matter. 
Heavy-quarks crossing through the QCD matter might interact with the medium and lose energy via various mechanisms. 
On one hand, the radiative energy loss is expected to be proportional to the QCD Casimir coupling factor between the hard parton and the gluons of the medium~\cite{bdmps-as}. Light-quarks should then lose less energy than gluons due to their color charge. 
In addition, due to their large mass, gluon-bremsstrahlung from heavy-quarks should be suppressed at small angles as a consequence of destructive quantum interferences~\cite{deadcone}. 
On the other hand, if heavy quarks hadronize in the medium or if they re-interact strongly with it, heavy-flavor hadrons should inherit the medium azimuthal anisotropy. 
Finally, if in-medium hadronization is the predominant mechanism at low momentum, the relative production of strange over non-strange charmed hadrons  should be enhanced~\cite{rafelski,he}. 
RHIC results show that heavy quarks lose energy in the medium, but a possible quark-mass hierarchy has not yet been elucidated.

The ALICE apparatus has the ability to detect, within others, heavy-flavor hadrons in various decay channels and in a wide kinematic range. 
The experimental setup and the strategy followed for these measurements are explained in Sec.~\ref{sec:detector}. 
The main detection modes studied are charmed mesons hadronic decays (mid-rapidity), and heavy-flavor decay electrons (mid-rapidity) and muons (forward-rapidity). 
The results in \pp~collisions at $\sqrts=7$~TeV and $2.76$~TeV are summarized in Sec.~\ref{sec:ppresults}. 
The measurements in \PbPb~collisions at $\sqrtsnn=2.76$~TeV are discussed and compared to other measurements and model calculations in Sec.~\ref{sec:PbPbresults}.

\section{Heavy-flavor measurements in ALICE}

\label{sec:detector}
The ALICE detector~\cite{ALICEJinst} consists of a central barrel ($|\eta|<0.9$), a muon spectrometer ($-4.0 <\eta < -2.5$) and a set of detectors for trigger and event characterization purposes. 
The central barrel is equipped with a solenoid magnet delivering a magnetic field of up to 0.5~T, 
and is in charge of reconstructing and identifying charged particles, photons and jets.
The Inner Tracking System (ITS) and the Time Projection Chamber (TPC) provides track reconstruction from very low ($\sim 100~\mev/c$) up to fairly high ($\sim 100~\gev/c$) transverse momentum with a momentum resolution better than $4\%$ for $\pt<20~\gev/c$. 
The good impact parameter (distance of closest approach of the track to the primary interaction vertex) resolution, better than 65~$\mu$m for $\pt>1~\gev/c$ in the bending plane in \PbPb~collisions, allows the reconstruction of secondary decay vertices, making possible the direct reconstruction of charmed hadrons and the measurement of beauty decay electrons. 
Charged hadrons are separated using the particle specific energy deposit (d$E$/d$x$) in the TPC and the timing in the Time Of Flight (TOF) detector. 
Electron identification (ID) is performed via the information provided by the TPC and TOF for $\pt \lesssim 6~\gev/c$, while the Transition Radiation Detector (TRD) and Electromagnetic Calorimeter (EMCAL) contribute to electron ID for $p>2$--$3~\gev/c$.
The muon spectrometer is equipped with a dipole magnet providing a magnetic field of up to 0.7~T, and is responsible for muon tracking and identification. From the vertex region to the outer part it is composed of a 10 interaction length passive absorber, a beam shield, five stations of cathode pad chambers (tracking stations), an iron wall, and two stations of resistive plate chambers (trigger stations). 
The VZERO detector, comprising two scintillator hodoscopes and located in the forward and backward regions, is involved in fast triggering and centrality determination. 
The experiment is completed by a set of global and central detectors that are not used in the measurements presented here.

\label{sec:hfeStrategy} 

Heavy-flavor decay electrons, ${\rm D/B} \longrightarrow {\rm e} +X$, were studied in $|\eta|<0.8$~\cite{ALICEHFepp7TeV}. Their spectrum was obtained by subtracting (statistically) a cocktail of the non-heavy-flavor decay sources to the inclusive electron spectrum. The cocktail contained the contributions of:  Dalitz decays of light neutral mesons ($\pi^0$, $\eta$, $\omega$, $\eta^{\prime}$, $\phi$), dilepton decays of vector mesons ($\rho$, $\omega$, $\phi$) and quarkonia ($\Jpsi$, $\Upsilon$), $\gamma$-conversions in the material and direct radiation. It was based on the measured $\pi^0$, $\Jpsi$, $\Upsilon$ spectra (the last only in \pp) and direct $\gamma$ from NLO calculations. 
In \PbPb~collisions, the contributions of Dalitz decays and $\gamma$-conversions were evaluated with an invariant mass analysis.

Beauty decay electrons, ${\rm B} \longrightarrow {\rm e} +X$ were separated with two independent methods in \pp~collisions.  One is based on the lifetime of beauty hadrons ($c\tau\sim500~\mu$m), and explores the separation of their decay electrons from the primary interaction vertex (IP) by cutting on their impact parameter distribution~\cite{ALICEHFBepp7TeV}. The other method consists on studying the azimuthal correlations of electrons and charged hadrons, particularly the near side region, whose width is larger for beauty than charm hadron decays due to the decay kinematics.

\label{sec:hfmStrategy}

Muons were reconstructed in $-4.0 <\eta < -2.5$ with the muon tracking chambers and identified by requiring the tracks to traverse the muon iron filter and reach the muon trigger chambers. 
The remaining background contribution is originated by primary light hadrons decays, mainly $\pi$ and K, and is evaluated with Monte Carlo simulations in \pp~collisions and extrapolating the ALICE mid-rapidity $\pi$ and K measurements in \PbPb~collisions~\cite{ALICEHFmpp7TeV,ALICEHFmRaa}.

\label{sec:DStrategy}

Charmed mesons were reconstructed through their hadronic decays: ${\rm D}^0 ({\bar{\rm D}^{0}}) \to {\rm K}^{\mp}\pi^{\pm}$, ${\rm D}^{\pm} \to {\rm K}^{\mp} \pi^{\pm} \pi^{\pm}$, ${\rm D}^{*\pm}(2010) \to {\rm D}^0 ({\bar{\rm D}^0}) \pi^{\pm}$ and ${\rm D}_{\rm s}^{\pm} \to \phi \pi^{\pm} \to {\rm K}^{\pm} {\rm K}^{\mp} \pi^{\pm}$. 
The analysis strategy exploits the apparatus ability to reconstruct the secondary vertices of the D meson decays ($\Dzero$ in the case of $\Dstar$). The large combinatorial background is reduced applying, $\pt$~and meson dependent, constraints on their decay topology and identifying their decay products. 
The prompt charm (${\rm c}\to{\rm D}$) spectrum were determined in $|y|<0.5$ after subtracting the feed-down contribution (${\rm b}\to{\rm D}$), based on FONLL~\cite{fonll} pQCD calculations and the ${\rm B}\to{\rm D}$ EvtGen~\cite{evtgen} decay kinematics~\cite{ALICEDpp7TeV,ALICEDpp276TeV,ALICEDspp7TeV,ALICEDRaa}.


\vspace{-8pt}
\section{Production in \pp~collisions at $\sqrts=7$~TeV and $\sqrts=2.76$~TeV}
\label{sec:ppresults}

The $\pt$-differential production cross sections of prompt charmed mesons ($\Dzero$, $\Dplus$, $\Dstar$, $\Ds$), heavy-flavor decay electrons, heavy-flavor decay muons, and beauty-decay electrons in \pp~collisions at $\sqrts=7$~TeV were reported in~\cite{ALICEDpp7TeV,ALICEDspp7TeV,ALICEHFepp7TeV,ALICEHFmpp7TeV,ALICEHFBepp7TeV}. 
The respective measurements of prompt charmed mesons ($\Dzero$, $\Dplus$, $\Dstar$) and heavy-flavor decay muons in \pp~collisions at $\sqrts=2.76$~TeV were presented in~\cite{ALICEDpp276TeV,ALICEHFmRaa}. 
The results are well described by pQCD predictions. 
FONLL~\cite{fonll} calculations describe reasonably well the differential cross sections. 
GM-VFNS~\cite{gmvfns} describes prompt D meson measurements in the $\pt$ range where the model is valid ($>3~\gev/c$). 
The total ${\rm c}\bar{\rm c}$ and mid-rapidity ${\rm b}\bar{\rm b}$ production cross sections were evaluated extrapolating the $\pt$-differential cross sections with pQCD calculations~\cite{ALICEDpp276TeV,ALICEHFBepp7TeV}. 
Their $\sqrts$--evolution are well described by NLO pQCD calculations (MNR~\cite{mnr}).


The limited statistics of the \pp~data sample collected on March 2011 at $\sqrts=2.76$~TeV prevents the usage of these measurements as reference for the \PbPb~studies at $\sqrtsnn=2.76$~TeV for prompt charmed hadrons and heavy-flavor decay electrons. 
For these probes, the \pp~reference was obtained by a pQCD-driven $\sqrt{s}$-scaling of the \pp~cross sections at $\sqrts=7$~TeV. The scaling factor was evaluated as the ratio of the FONLL calculations~\cite{fonll} at these two energies, and its uncertainty was determined by the envelope of the ratios while varying the calculation parameters (factorization and renormalization scales) and the heavy-quark masses~\cite{scaling}. The procedure was validated by comparing the results to the measurement of prompt charmed mesons in \pp~collisions at $2.76$~TeV and to CDF Tevatron data $1.96$~TeV, see~\cite{ALICEDpp276TeV,scaling} and references therein. 
Finally, for the high-$\pt$ range of the heavy-flavor decay electrons ($\pt>8~\gev/c$) and the highest $\pt$-bin of the prompt charmed mesons ($\pt>16~\gev/c$), no \pp~cross section could be scaled. There, the FONLL calculation and a $\pt$-extrapolation of the cross sections were respectively evaluated.

\vspace{-8pt}
\section{Suppression and azimuthal anisotropy in \PbPb~collisions at $\sqrtsnn=2.76$~TeV}
\label{sec:PbPbresults}

The relative production of particles in heavy-ion collisions with respect to that in nucleon-nucleon interactions is commonly studied evaluating the nuclear modification factor, defined as  
\begin{equation}
\RAA(\pt) =  1 / \langle \TAA \rangle \cdot  \left(  {\rm d} N_{\rm AA}/{\rm d}\pt  \big/ {\rm d}\sigma_{\rm pp}/{\rm d}\pt   \right)\, , 
\end{equation}
the ratio between the yield in AA collisions (${\rm d} N_{\rm AA}/{\rm d}\pt$), and the \pp~cross section (${\rm d}\sigma_{\rm pp}/{\rm d}\pt $) normalized by the average nuclear overlap function ($\langle \TAA \rangle$). $\TAA$ is the convolution of the colliding ions nuclear density profile and is evaluated in the Glauber model. 

\begin{figure}[!htbp]
\begin{center}
 \includegraphics[width=0.46\textwidth]{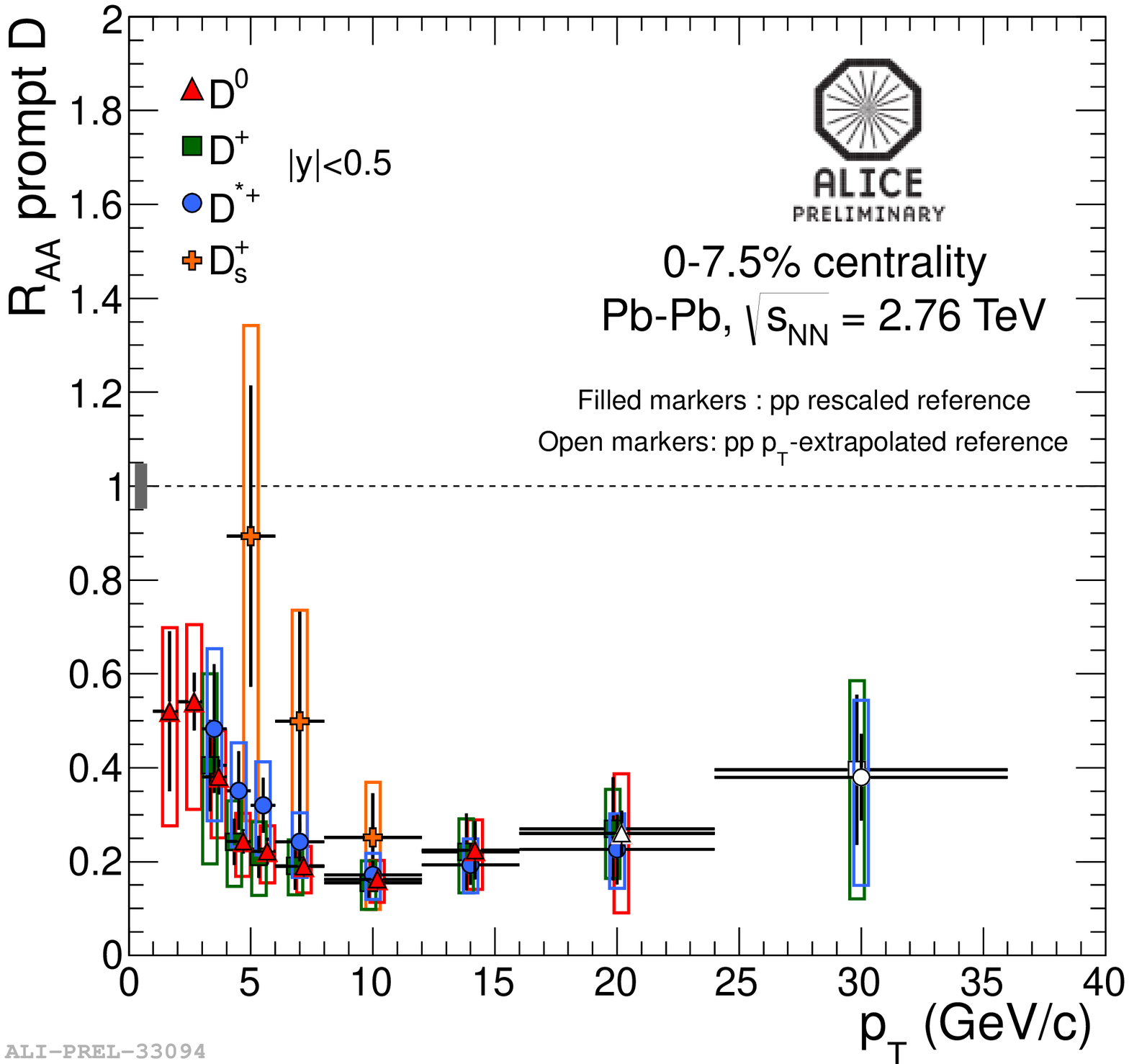}
 \includegraphics[width=0.46\textwidth]{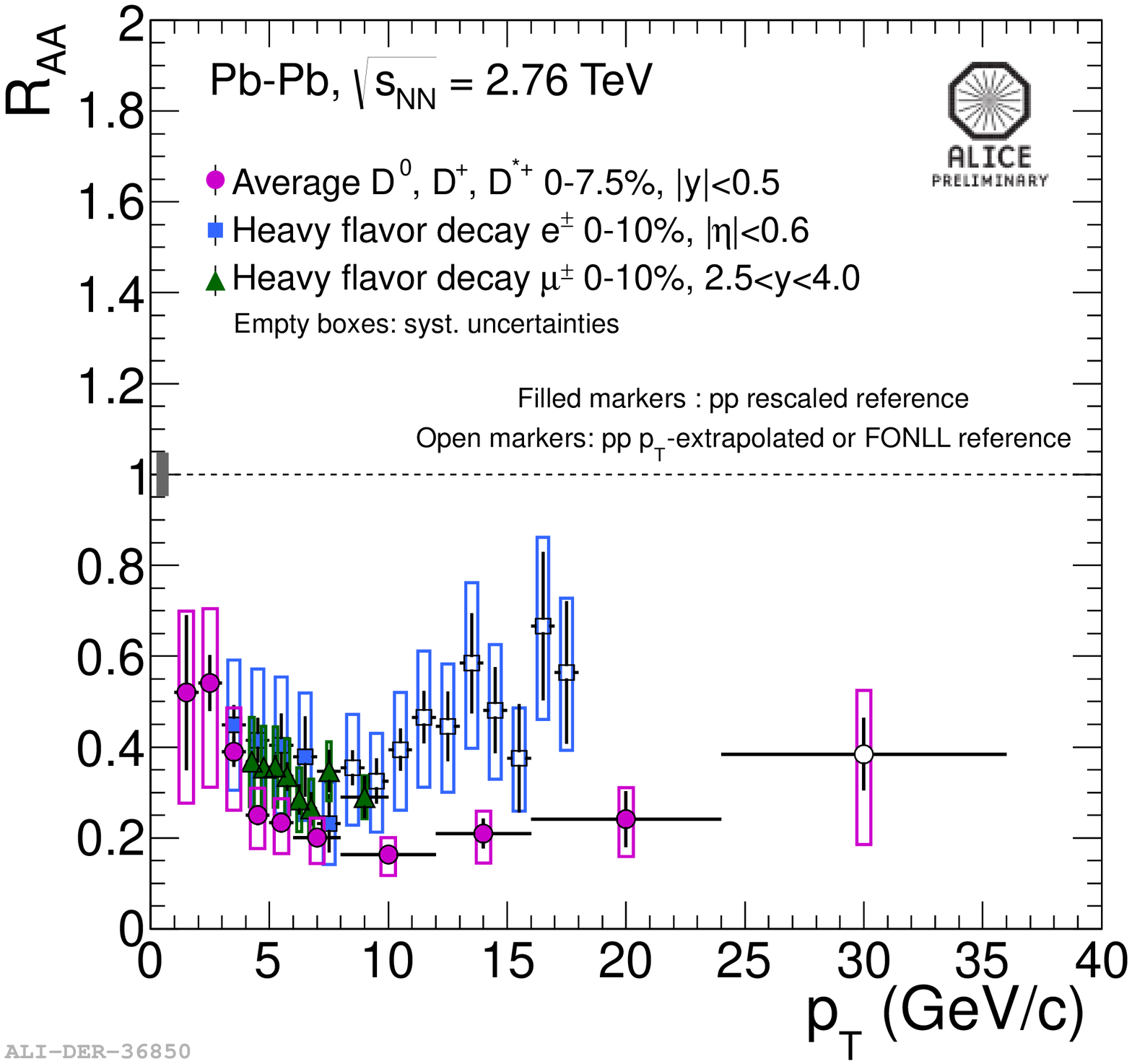}
 \caption{
 	$\RAA(\pt)$ in the most central \PbPb~collisions at $\sqrtsnn=2.76$~TeV. 
	Left: Prompt $\Dzero$, $\Dplus$, $\Dstar$ and $\Ds$ mesons.
 	Right: Heavy-flavor decay electrons (blue squares), heavy-flavor decay muons (green triangles)~\cite{ALICEHFmRaa} and the average of prompt $\Dzero$, $\Dplus$ and $\Dstar$ (violet circles). 
	\label{fig:HFRaaVsPt}
 }
\end{center}
\end{figure}

Heavy-flavor decay muon and prompt $\Dzero$, $\Dplus$ and $\Dstar$ meson $\RAA$ $\pt$ and centrality dependence were first measured with the 2010 \PbPb~data sample collected with a minimum bias trigger~\cite{ALICEDRaa,ALICEHFmRaa}. 
Heavy-flavor decay muon $\RAA(\pt)$ in the 0--10\% most central class is shown Fig.~\ref{fig:HFRaaVsPt}~(right). 
The 2011 \PbPb~data sample, collected with the combination of a minimum bias trigger and a centrality trigger based on the VZERO scintillators response, allowed to roughly collect a factor of 10 more statistics in the 0--7.5\% centrality class than in the 2010 data sample, and triplicate the statistics it in the 10--50\% centrality class. In addition, the EMCAL triggers allowed to enhance the electron sample.

\label{sec:RAAresults}

Prompt D meson $\RAA(\pt)$ was measured in the 0--7.5\% centrality class with the 2011 data sample up to $36~\gev/c$, see Fig.~\ref{fig:HFRaaVsPt}~(left)~\cite{grelli,innocenti}. 
$\Dzero$, $\Dplus$ and $\Dstar$ meson $\RAA(\pt)$ agree within their uncertainties  and show a suppression of up to a factor of 5 at $\pt\sim10~\gev/c$. 
The first measurement of $\Ds$ $\RAA$ in heavy-ion collisions is also shown. In the 8--12 $\gev/c$ $\pt$ bin, the $\Ds$ $\RAA$ is compatible with that of non-strange charmed mesons. At lower $\pt$, $\Ds$ $\RAA$ seems to increase, but the current statistical and systematic uncertainties preclude any significant comparison to the $\Dzero$, $\Dplus$ and $\Dstar$ meson $\RAA(\pt)$. 
The weighted average of $\Dzero$, $\Dplus$ and $\Dstar$ meson $\RAA$ is compared in Fig.~\ref{fig:HFRaaVsPt}~(right) with heavy-flavor decay electron and muon $\RAA$ in the most central collisions. 
Heavy-flavor decay electron $\RAA(\pt)$ was measured in the 0-10\% centrality class using the EMCAL triggered 2011 \PbPb~data sample~\cite{sakai}. It exhibits a suppression for $3<\pt<18~\gev/c$, that amounts to a factor of 1.5--3 for $3<\pt<10~\gev/c$.
Heavy flavor decay leptons and prompt D mesons present a similar suppression magnitude and $\pt$ dependence in the most central collisions, if we consider the B/D hadron to lepton decay kinematics.

\label{sec:discussion}

\begin{figure}[htbp]
\begin{center}
 \includegraphics[width=0.47\textwidth]{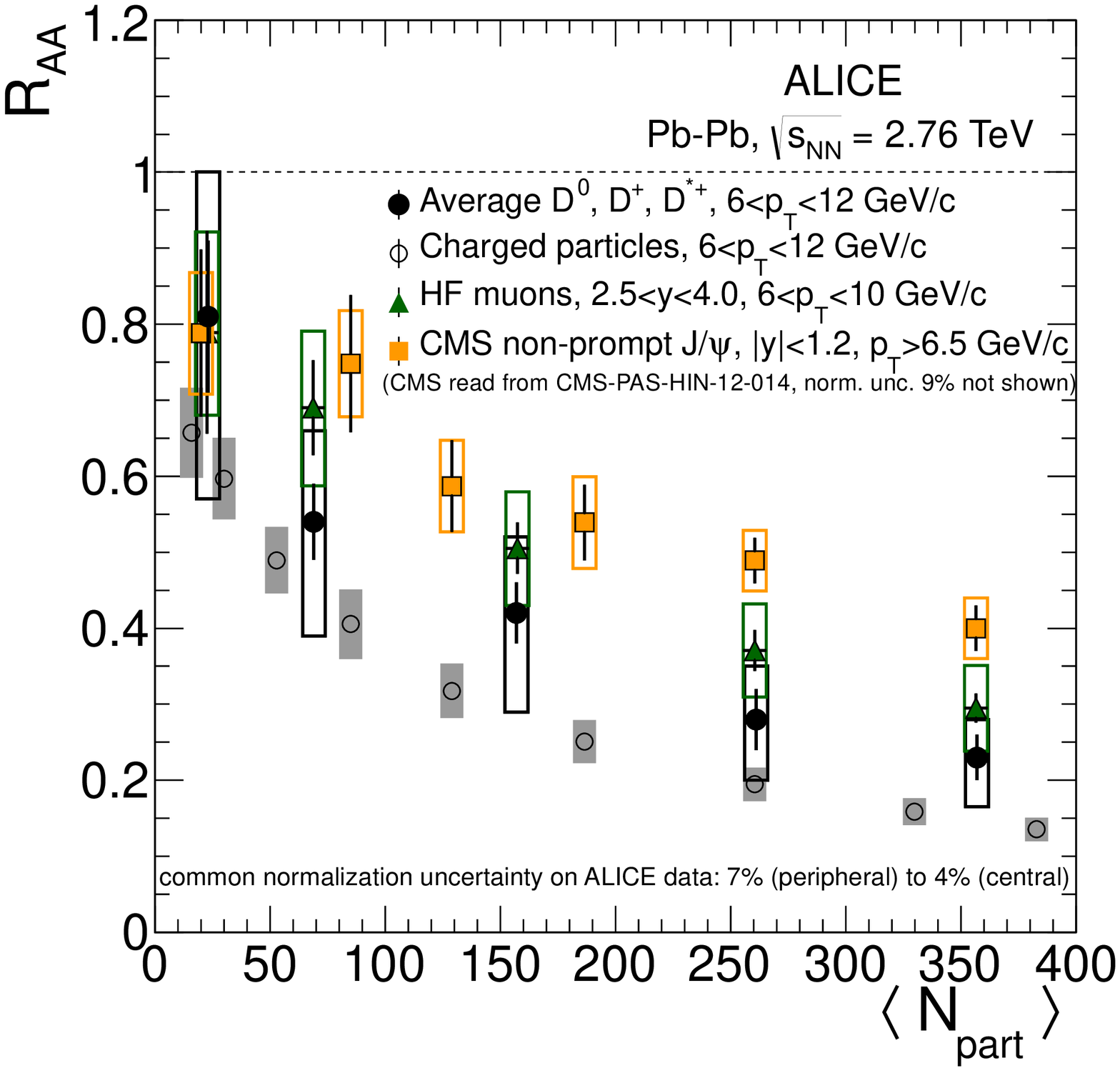}
  \includegraphics[width=0.475\textwidth,height=0.465\textwidth]{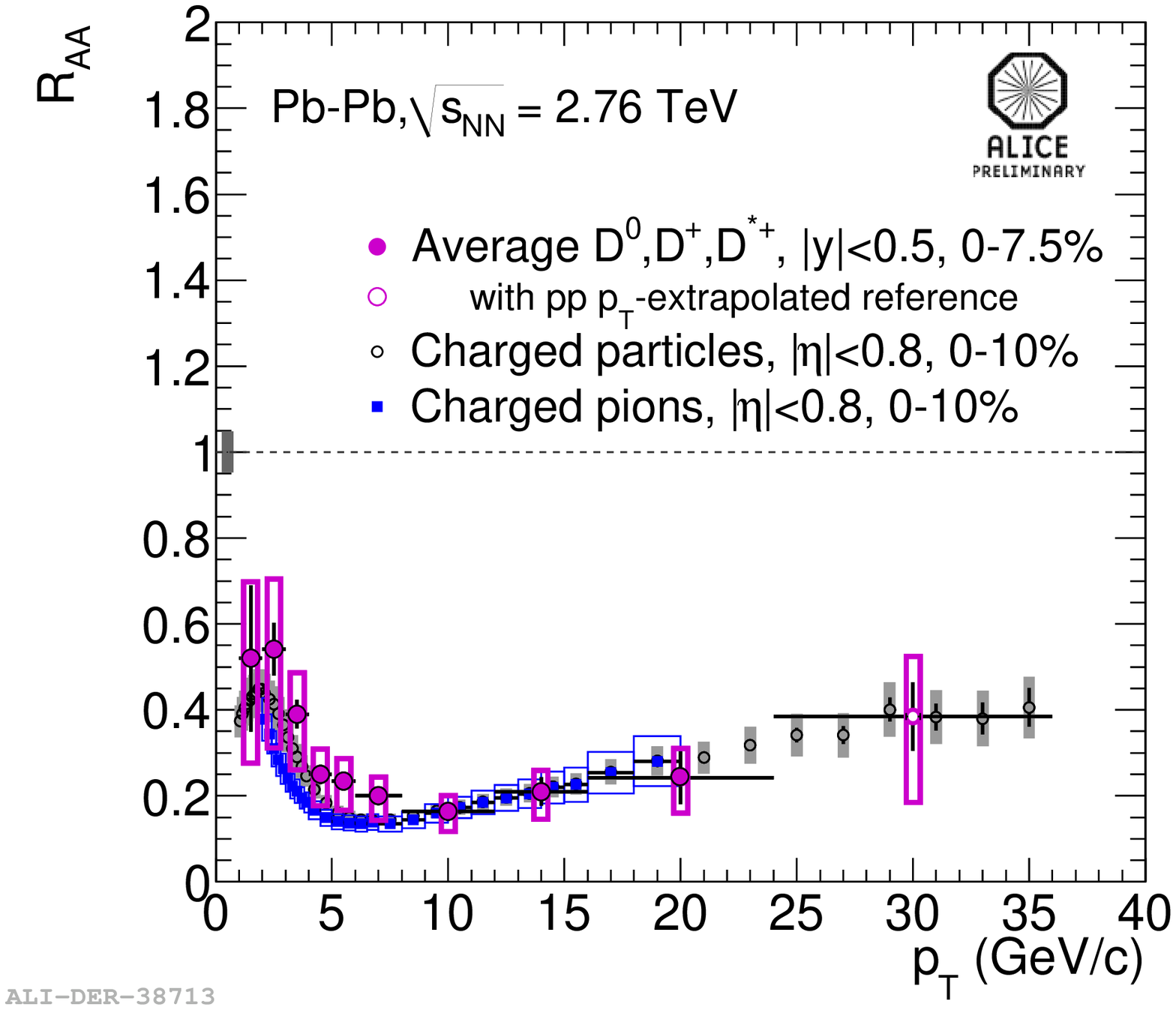} 
 \caption{ 
	 Left: $\RAA$ centrality dependence for $\pt>6~\gev/c$ of the heavy-flavor decay muons (green triangles), the average prompt D mesons (black filled circles), charged particles (empty black circles) and non-prompt $\Jpsi$ from CMS~\cite{CMSquarkonia,CMSHIN1214}.
	Right: 
	$\RAA(\pt)$ of the prompt D-meson average (violet circles), charged particles (black circles), and charged pions (blue squares) in the most central class~\cite{ALICENchargedRaa}. 
 \label{fig:RaaVsChargedPart}
 \label{fig:RaaDNchPiPt}
 }
\end{center}
\end{figure}

Heavy-flavor suppression centrality dependence is shown in Fig.~\ref{fig:RaaVsChargedPart}~(left) for $\pt>6~\gev/c$. 
Prompt D mesons and heavy-flavor decay muons present a similar $\RAA$ magnitude and centrality dependence, although the measurements correspond to different rapidity ranges. 
Charged particles' trend~\cite{ALICENchargedRaa} also exhibits a resemblance to that of D mesons. 
Non-prompt $\Jpsi$ (from CMS) data~\cite{CMSquarkonia,CMSHIN1214} is consistent with that of heavy-flavor decay muons, mainly populated of beauty decays in this $\pt$ range. 
In addition, the $\RAA(\pt)$ of D mesons, charged particles and pions in the most central class, see Fig.~\ref{fig:RaaDNchPiPt}~(right), are compatible within the present uncertainties. 
In conclusion, the measurements illustrate a consistent picture, with an indication of $\RAA$ of non-prompt $\Jpsi$ larger than that of prompt D mesons in the most central collisions, 
although their current precision prevents to conclude with respect to the expected color charge and mass hierarchy of parton energy loss. 

\begin{figure}[htbp]
\begin{center}
 \includegraphics[width=0.47\textwidth]{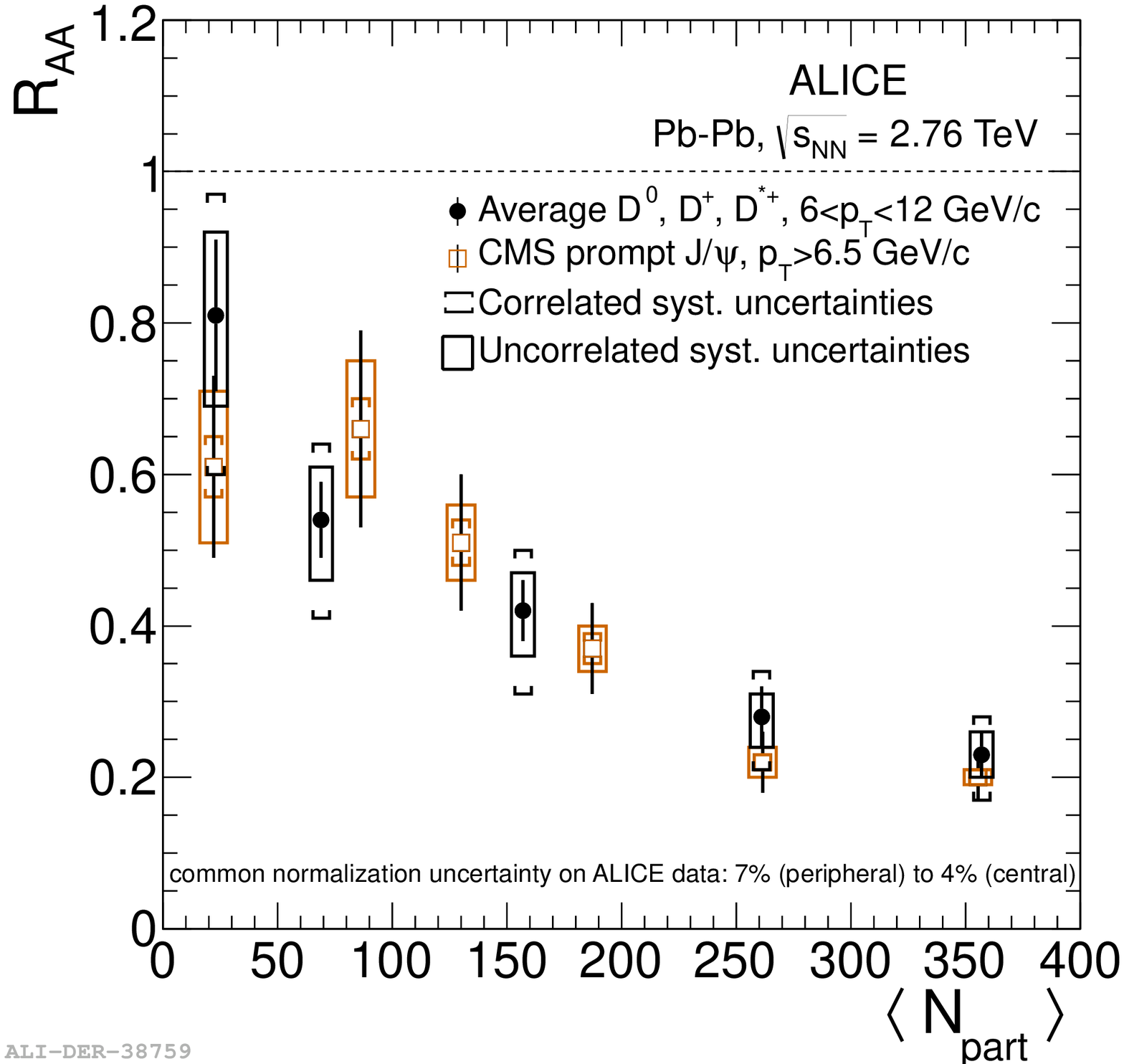}
 \includegraphics[width=0.47\textwidth]{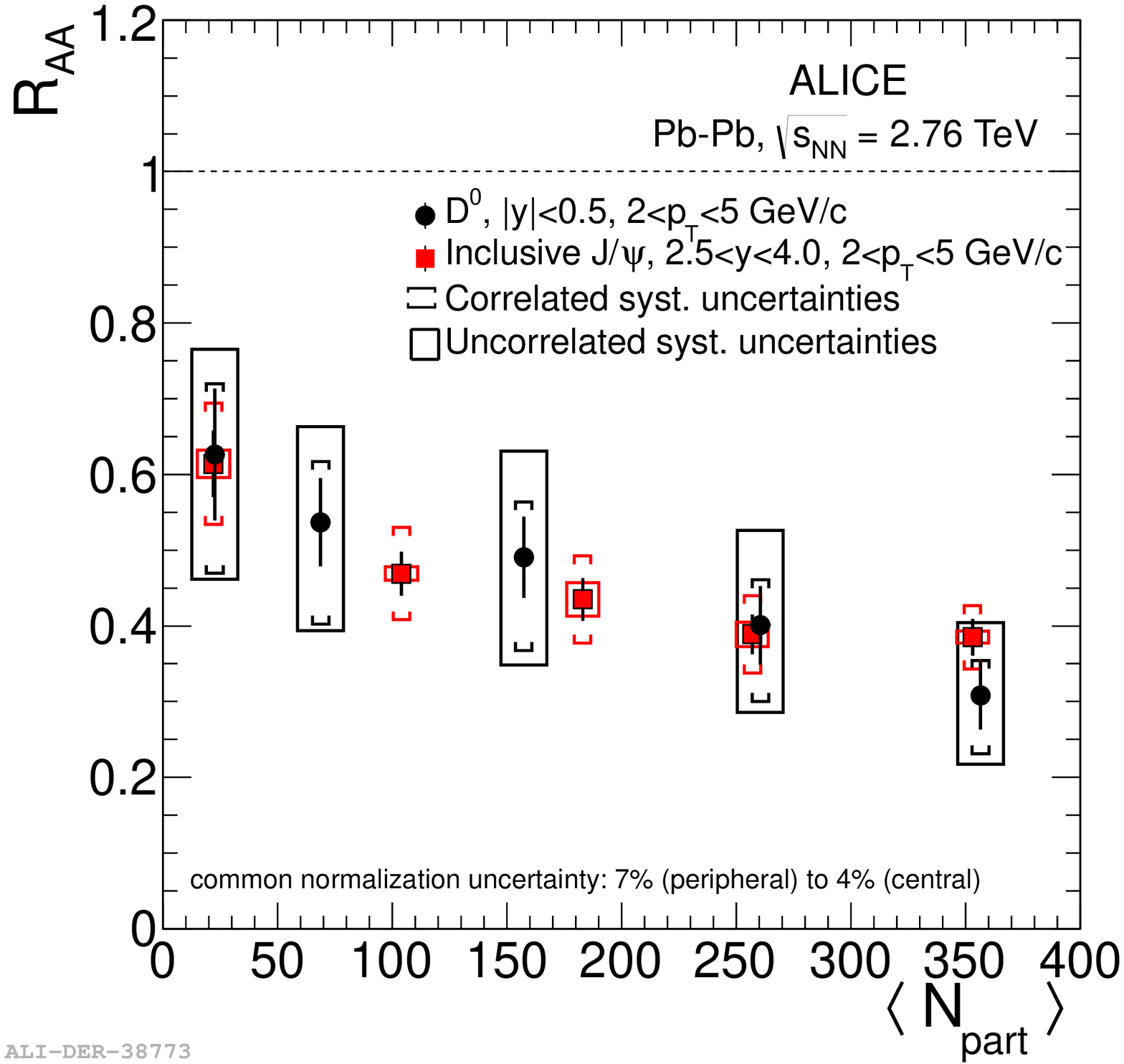}
 \caption{
	Left: $\RAA$ centrality dependence for $\pt>6~\gev/c$ of the prompt D-meson average (black filled circles), and prompt $\Jpsi$ from CMS~\cite{CMSquarkonia}. 
	Right: $\RAA$ centrality dependence for $2<\pt<5~\gev/c$ of the prompt D-meson average (black filled circles) in $|y|<0.5$, and inclusive $\Jpsi$ in $2.5<y<4.0$. 
 \label{fig:RaaPromptJpsiD}
 \label{fig:RaaInclJpsiD}
 }
\end{center}
\end{figure}

Open-charm production was proposed as reference for the studies of $\Jpsi$ production in a hot and dense QCD medium. It might bring some insight to distinguish the effects of the QCD medium on c quarks and ${\rm c}\bar{\rm c}$ bound states. 
The $\RAA$ centrality dependence of prompt D mesons and prompt $\Jpsi$ from CMS for $\pt>6~\gev/c$ at mid-rapidity are compared in Fig.~\ref{fig:RaaPromptJpsiD}~(left). Figure~\ref{fig:RaaPromptJpsiD}~(right) presents the $\RAA$ centrality dependence of prompt D mesons at $|y|<0.5$ and inclusive $\Jpsi$ at $2.5<y<4.0$ for $2<\pt<5~\gev/c$. The results exhibit a similar magnitude and centrality trend both at low and high $\pt$. 

\label{sec:V2results}

The initial geometrical anisotropy of non-central heavy-ion collisions is reflected in a momentum anisotropy of the produced particles caused by the different pressure gradients. It can be quantified by a Fourier decomposition of the azimuthal distribution of particles, $\varphi$, with respect to the event plane, defined by the beam direction and the impact parameter, $\psi$, 
\begin{equation}
{\rm d}N / {\rm d}\varphi =  N_0 / (2\pi) \left( 1+ 2v_1 \cos  \left(\varphi-\psi_1 \right) + 2v_2 \cos  \left[ 2 \left(\varphi-\psi_2 \right) \right]  + \dots \right)
.
\end{equation}
In particular, the measurement of the second Fourier coefficient, $\vtwo$, and $\RAA$ should bring some insight on the degree of thermalization of the medium and the path length dependence of the energy loss. The are multiple methods to measure $\vtwo$. For the heavy-flavor studies reported here we exploited three methods: the multi-particle correlations, fitting the azimuthal particle distributions, and the direct comparison of the yields produced in-plane and out-of-plane. 

\begin{figure}[!htbp]
\begin{center}
 \includegraphics[width=0.49\textwidth]{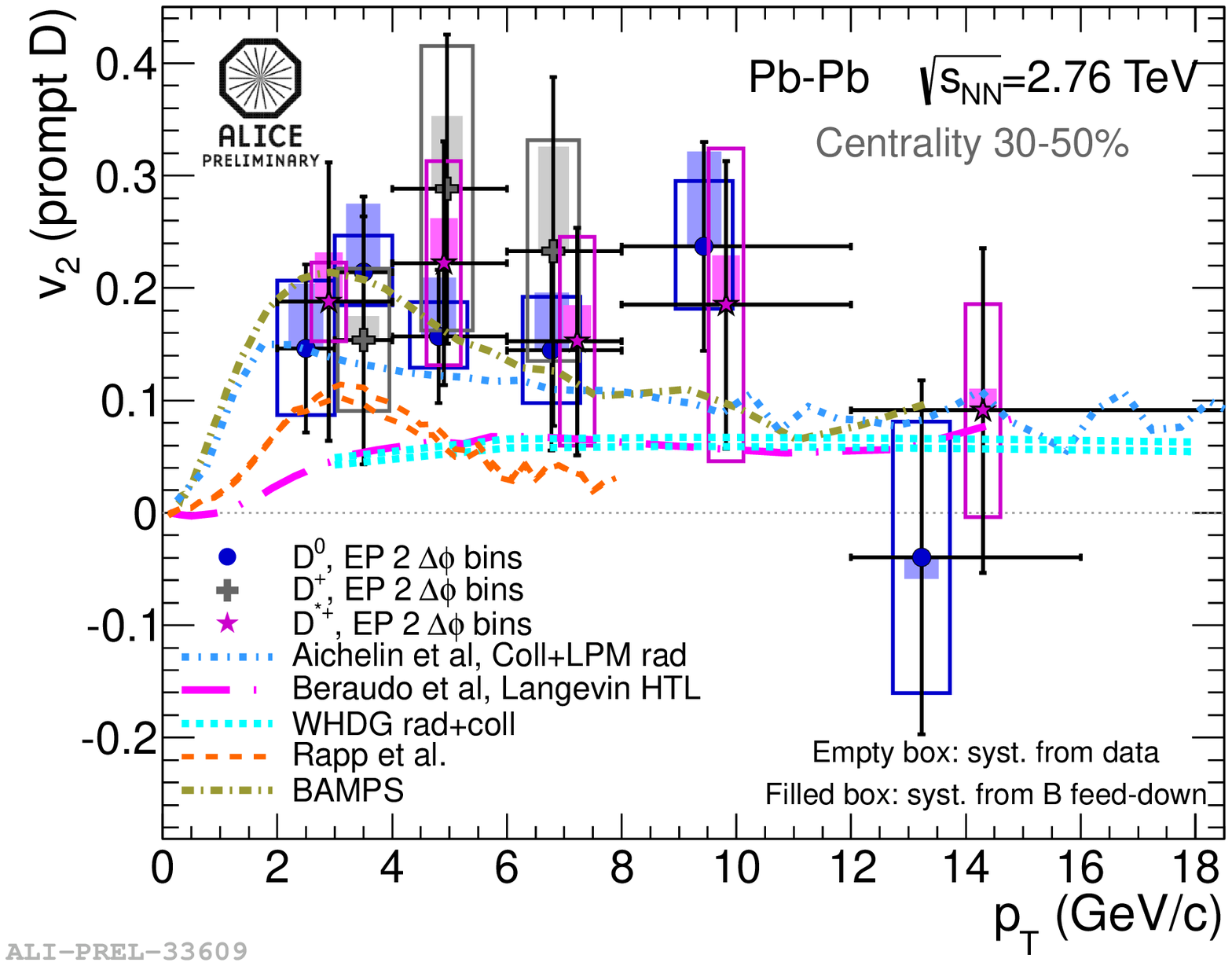} 
 \includegraphics[width=0.445\textwidth]{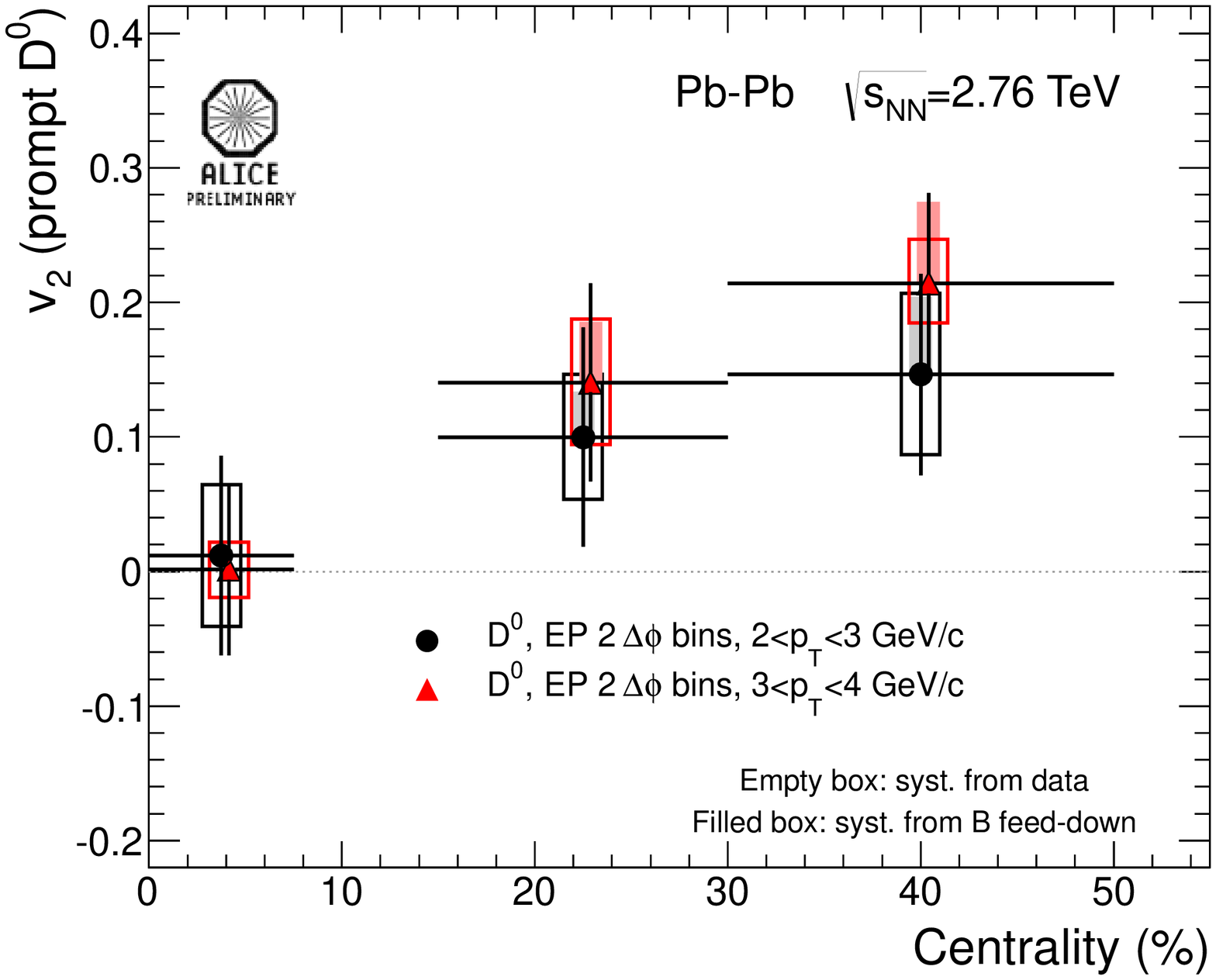}
 \caption{
 Left: prompt $\Dzero$, $\Dplus$ and $\Dstar$ $\vtwo$ in the 30--50\% centrality class versus transverse momentum compared to theoretical calculations~\cite{gossiaux,beraudo,whdg2011,rapp,bamps}. 
 Right: prompt $\Dzero$ $\vtwo$ versus centrality in two $\pt$ ranges. 
 \label{fig:DRaaEPandV2VsCent}
 }
\end{center}
\end{figure}

\begin{figure}[htbp]
\begin{center}
 \includegraphics[width=0.875\textwidth]{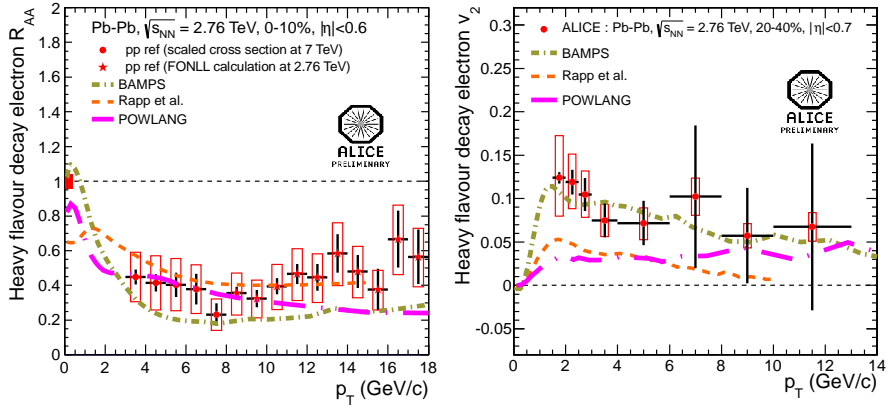}
 \caption{
 Heavy-flavor decay electron $\RAA$ in the 0--10\% most central class (left), and $\vtwo$ in the 20--40\% centrality class (right), compared to theoretical calculations: BAMPS~\cite{bamps}, Rapp et al~\cite{rapp}, POWLANG~\cite{beraudo}.
 \label{fig:HFeRaaV2Models}
 }
\end{center}
\end{figure}

The 2011 data sample allowed us to measure the heavy-flavor decay electron $\vtwo$ for $1.5<\pt<13~\gev/c$ in the 20--40\% centrality class~\cite{sakai} and prompt D meson $\vtwo$ for $2<\pt<18~\gev/c$ in the 30--50\% centrality class~\cite{caffarri}. 
The results of $\Dzero$, $\Dplus$ and $\Dstar$ $\vtwo(\pt)$ are consistent within statistical uncertainties, see Fig.~\ref{fig:DRaaEPandV2VsCent}~(left). 
At low $\pt$, $\vtwo>0$ with a $3\sigma$ effect, both for heavy-flavor decay electrons in the $2<\pt<3~\gev/c$ range, see Fig.~\ref{fig:HFeRaaV2Models}~(right), and prompt D mesons in the $2<\pt<6~\gev/c$ range. 
Prompt $\Dzero$ $\vtwo(\pt)$ was also measured in the 0--7.5\% and 15--30\% centrality classes. 
The results, see Fig.~\ref{fig:DRaaEPandV2VsCent}~(right), indicate a centrality dependence of $\vtwo$. 
Finally, prompt $\Dzero$ $\RAA(\pt)$ was evaluated in the in-plane and out-of-plane azimuthal regions for the 30--50\% centrality class~\cite{caffarri}. A larger suppression is measured in the out-of-plane azimuthal range. 
At low $\pt$, both the elliptic flow and the path length dependence of heavy-quark energy loss could explain the results, while at high $\pt$ the last seems to be the most probable scenario.

\label{sec:theory}

\begin{figure}[htbp]
\begin{center}
 \includegraphics[width=0.45\textwidth]{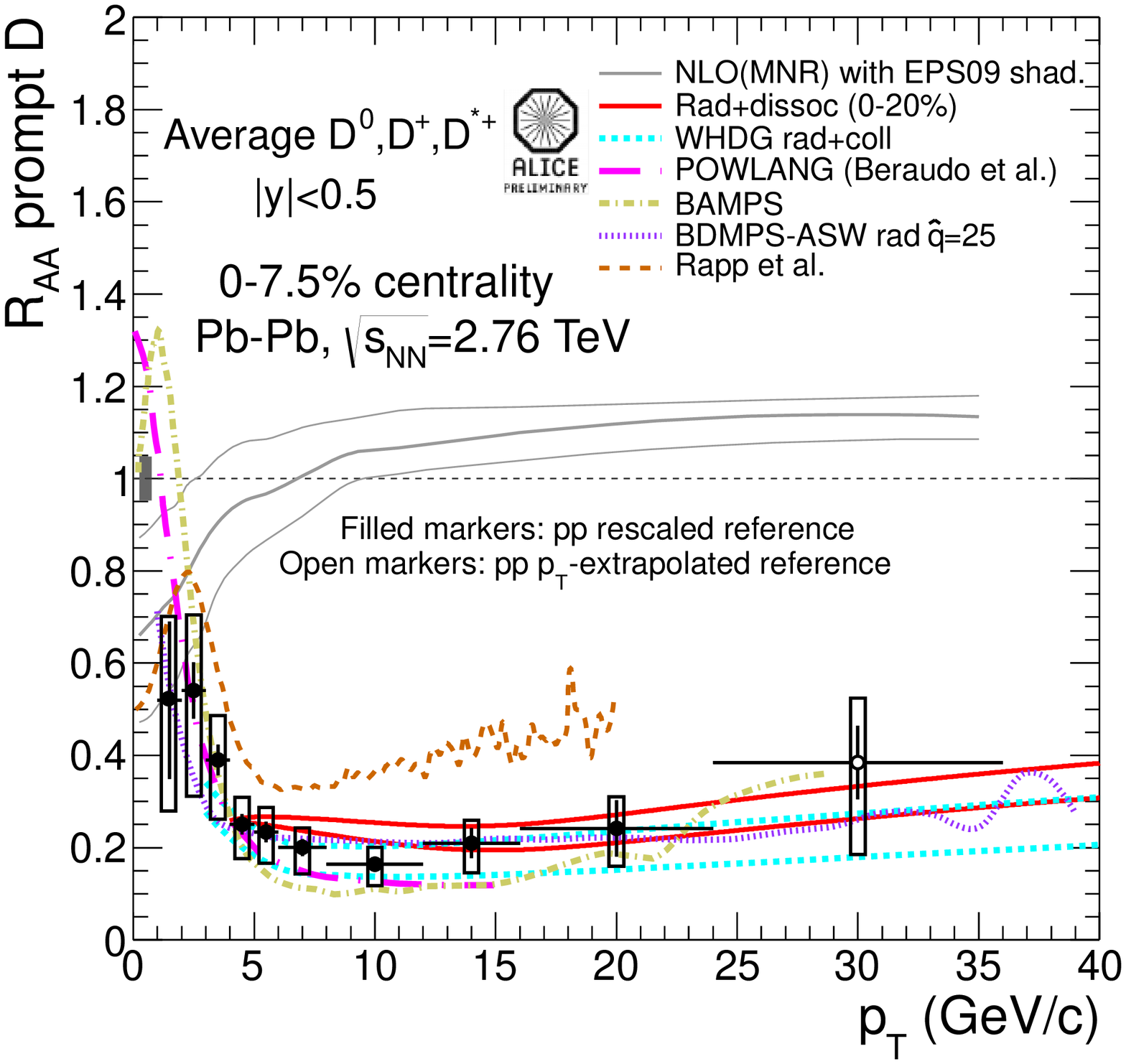}
 \includegraphics[width=0.425\textwidth,height=0.415\textwidth]{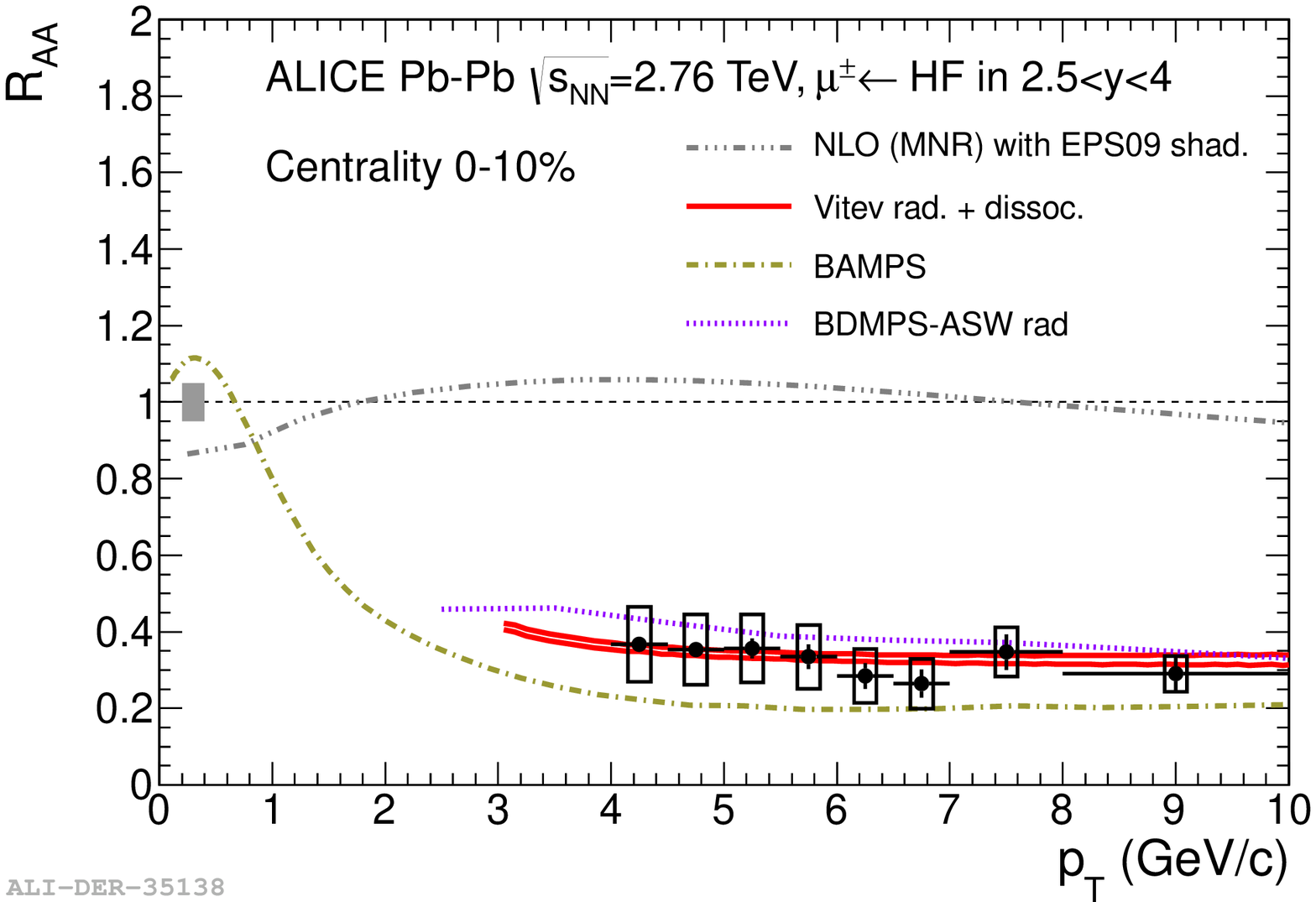}
 \caption{
 	Prompt average D meson (left) and heavy-flavor decay muon (right) $\RAA(\pt)$ in the most central collisions compared to NLO MNR calculations~\cite{mnr} with EPS09 shadowing parameterizations~\cite{eps09} and various in-medium models: BDMPS-ASW~\cite{bdmps-as}, WHDG~\cite{whdg2011}, BAMPS~\cite{bamps}, Rapp et al.~\cite{rapp}, POWLANG~\cite{beraudo}, Vitev et al.~\cite{vitev,vitevjet}.
 	\label{fig:RaaVsPtModels}
 }
\end{center}
\end{figure}

D-meson and heavy-flavor muon $\RAA(\pt)$ in the most central class are compared to NLO MNR calculations~\cite{mnr} with EPS09 shadowing parameterizations~\cite{eps09} in Fig.~\ref{fig:RaaVsPtModels}. Heavy-flavor suppression can not be explained with only nuclear shadowing for $\pt>4~\gev/c$. 
Several models based on parton energy loss calculate the heavy-flavor nuclear modification factor and/or the azimuthal anisotropy~\cite{bdmps-as,vitev,vitevjet,whdg2011,beraudo,gossiaux,bamps,rapp,cujet}. They are compared in Figs.~\ref{fig:DRaaEPandV2VsCent},~\ref{fig:RaaVsPtModels} and \ref{fig:HFeRaaV2Models} to the $\RAA$ and $\vtwo$ results. 
Those calculating radiative energy loss with heavy-quark in-medium dissociation (Vitev et al.)~\cite{vitev},
or both radiative and collisional energy loss (WHDG, CUJET)~\cite{whdg2011,cujet}, 
describe reasonably well heavy and light-flavor suppression~\cite{ALICEDRaa}. 
However, the models considering radiative energy loss without an hydrodynamical expansion of the medium do underestimate $\vtwo$ (POWLANG, WHDG)~\cite{beraudo,whdg2011}.
On the other hand, the heavy-quark transport based models: 
the one with collisional energy loss in a expanding medium (BAMPS)~\cite{bamps}, 
and the one with in-medium resonance scattering and coalescence (Rapp et al.)~\cite{rapp,he}, 
seem to respectively overestimate and underestimate heavy-flavor suppression. 
To conclude, the simultaneous description of heavy-flavor suppression and azimuthal anisotropy is challenging.  

\vspace{-10pt}
\section{Summary}

Heavy-flavor production has been studied in \pp~and \PbPb~collisions with the ALICE detector at the LHC. 
The differential production cross section of heavy-flavor hadrons (heavy-flavor decay leptons, beauty-decay leptons and prompt D mesons) in \pp~collisions is well described by pQCD calculations. 
Their production in \PbPb~collisions at$\sqrtsnn=2.76$~TeV has been examined by computing their nuclear modification factor. 
The first measurement of the $\Ds$~meson $\RAA$ has been presented. 
Heavy-flavor hadrons are suppressed in the most central \PbPb~interactions. 
The suppression is similar for the three studied decay channels (heavy-flavor electrons, heavy-flavor muons, prompt D mesons), and amounts up to a factor of 3--5 for $\pt \sim 8$--$10~\gev/c$. 
NLO MNR calculations with EPS09 shadowing parameterization alone can not explain this suppression for $\pt>4~\gev/c$. 
The forthcoming p--Pb run in 2013 will allow to better characterize the cold nuclear matter influence. 
The azimuthal anisotropy has been characterized measuring $\vtwo$. 
It has been observed that $\vtwo>0$ with a $3\sigma$ effect for both heavy-flavor decay electrons and prompt D mesons in the semi-central collisions for $2<\pt<3~\gev/c$ and $2<\pt<6~\gev/c$ respectively. 
The values of prompt $\Dzero$ $\vtwo$ seem to decrease in more central collisions. 
Several models based on parton energy loss calculate $\RAA$ and/or $\vtwo$, but the simultaneous description of heavy-flavor hadron $\RAA$ and $\vtwo$ is challenging.

\vspace{-10pt}
\section*{References}


\end{document}